\documentclass[prl,aps,twocolumn,floats]{revtex4}

\usepackage{graphicx}
\usepackage{amsmath}
\usepackage{citesort}
\usepackage{dcolumn}

\begin{document}

\title{
{\rm\small\hfill (Chem. Phys. Lett., accepted)}\\
Robustness of ``cut and splice'' genetic algorithms\\ 
in the structural optimization of atomic clusters}

\author{Vladimir A. Froltsov and Karsten Reuter}

\affiliation{Fritz-Haber-Institut der Max-Planck-Gesellschaft,
Faradayweg 4-6, D-14195 Berlin, Germany}

\received{25th February 2009}

\begin{abstract}
We return to the geometry optimization problem of Lennard-Jones clusters to analyze the performance dependence of ``cut and splice'' genetic algorithms (GAs) on the employed population size. We generally find that admixing twinning mutation moves leads to an improved robustness of the algorithm efficiency with respect to this {\em a priori} unknown technical parameter. The resulting very stable performance of the corresponding mutation+mating GA implementation over a wide range of population sizes is an important feature when addressing unknown systems with computationally involved first-principles based GA sampling.
\end{abstract}

\maketitle

\section{Introduction}

Small clusters of less than a hundred atoms are an increasingly studied class of materials exhibiting unique optical, magnetic and chemical properties \cite{jena87}. Due to the intricate relationship between the geometric and electronic structure of clusters in this size range a reliable determination of the ground-state and energetically low-lying metastable geometric structures is a prerequisite. Two crucial aspects of the corresponding global geometry optimization problem are the reliable sampling of the high-dimensional potential-energy surface (PES) to identify the low-lying minima, and the accuracy with which this PES is provided \cite{wales03}. When aiming at a quantitative, material-specific theory the high computational cost to evaluate the PES with first-principles electronic structure methods adds another practical aspect, namely that the sampling strategies be as efficient as possible. High efficiency is in this respect often solely equated with a minimum number of PES evaluations required until the relevant structures are found. This neglects that sophisticated sampling algorithms are usually based on a manifold of technical parameters which sensitively determine this efficiency and which can mostly only be optimized through either a detailed knowledge of the PES topography or repeated sampling runs with different settings. Again, for numerically undemanding model PES the latter is not an issue, but instead done routinely. However, for costly first-principles sampling of unknown systems, where every single energy and force evaluation may take say several hours of CPU time, this is clearly not possible and it thus becomes equally important that the efficiency is robust, namely that the algorithm exhibits a similar performance as long as the chosen parameter settings are in a reasonable range.

\section{Theory}

With this motivation we return to an algorithm that is widely used for geometry optimization problems, the genetic algorithm (GA) \cite{goldberg89,hartke04}, and analyze its robustness with respect to a central technical parameter, the population size. In the context of structure optimization GAs explore the PES landscape through a sequence of trial structures that each time correspond to local PES minima. Such trial structures are generated by randomly modifying a given cluster structure in a so-called move and an ensuing local geometry optimization. The central idea of GAs is thereby to run several such trial structure sequences in parallel, exploiting this population to also create new trial structures by mating the different current configurations and thereby potentially combining information from disparate parts of the high-dimensional PES. Replacing old members of the population with newly obtained more stable candidates, the generation update procedure is repeated until the global optimized and other low-lying structures are found. A plethora of subtle modifications of this basic principle has been suggested, including parameter settings and move types tailored for specific applications (e.g. \cite{hartke04,deaven95,deaven96,wolf98,hartke99,johnston03}). With the goal of a robust general-purpose methodology with a minimum number of technical parameters we only return to these ramifications below, and focus instead here on a classic move type originally suggested by Deaven and Ho \cite{deaven95,deaven96}. Such ``cut and splice'' moves are generically applicable and by their very construction enable huge jumps in configuration space, thus promising a decent sampling. As the name suggests, a new trial structure is generated by suitably cutting an existing cluster geometry into two halves along an arbitrarily oriented plane and then splicing two halves together. This can either be done by recombining the two halves of the same cluster after rotation by a random angle, thus making it a so-called twinning mutation \cite{wolf98}, or by recombining the halves of different configurations in the current population, thereby creating a mating or crossover operation. While such mating moves appear most natural within the GA philosophy of mixing ``genetic'' information in the population, the beneficial effect of mutation moves in preventing premature convergence is well established \cite{wolf98,hartke99,johnston03}. This is hitherto primarily discussed with respect to an improved sampling efficiency, whereas our analysis below in fact demonstrates that it also has important bearings on the robustness of the approach.

With such move type fixed, the only algorithmic parameters left are the size of the population and the way how trial structures replace current members of the population if they are more ``fit'', i.e. energetically more stable. In a corresponding periodic population update a large number of new trial structures is first created by taking all possible pairings within the existing population and only doing the replacement step afterwards. Particularly for larger populations this bears the danger of creating a lot of new trial structures out of members that still have a low fitness. This can be circumvented by always allowing only a certain fraction of the most stable clusters in the population to mate, which comes at the price of adding with this fraction another unknown technical parameter with impact on the algorithm efficiency. Another alternative that gets away without this is a dynamic population update \cite{goldberg89}, which we indeed verified to be much more efficient than a periodic population update for all systems and settings discussed below. In this procedure new trial structures are still created by sequentially taking all possible pairings within the population, which then automatically includes twinning mutations in form of the self-pairings. However, in contrast to a periodic update the fitness of a new trial structure is immediately evaluated. If it is more stable than any of the two parent structures (one parent structure in case of the mutations), it directly replaces the highest energy member of the entire population, thereby quickly eliminating the least fit configurations. In the conceptual idea of a genetic algorithm genetic diversity is hereby enforced by rejecting trial structures that correspond to local PES minima that are already represented in the current population.

This leaves as the sole undetermined technical parameter of the present GA implementation the size of the population, the optimum value of which is {\em a priori} unknown for an unknown system. In order to arrive at a trend understanding of how much the algorithm efficiency depends on a judicious choice of this parameter we resort to the well-studied problem of the structural optimization of Lennard-Jones (LJ) clusters, i.e. clusters of atoms interacting with each other via the well-known LJ pair potential,
\begin{equation}
u_{\rm {LJ}}(r) \;=\; 
4\varepsilon_{\rm {LJ}} \left[\left( \frac{\sigma_{\rm {LJ}} }{r}\right) ^{12}-\left(\frac{
\sigma_{\rm {LJ}} }{r}\right) ^{6}\right] \quad ,
\label{u_LJ}
\end{equation}
where $r$ is the interparticle distance, $\sigma_{\rm {LJ}}$ is the effective particle diameter, $\varepsilon_{\rm {LJ}}$ sets the energy scale of the short-ranged soft core repulsion, and we use reduced units $\sigma_{\rm {LJ}}=\varepsilon_{\rm {LJ}}=1$ throughout. This simple model potential is chosen for a number of reasons: First, the global minima of LJ clusters in the targeted size range of up to 100 atoms are all well established \cite{wales97,cambridge_database}, thus providing a well-defined criterion for the efficiency of a sampling run, namely the number of trial moves $N$ until this global minimum is found. Second, the ease with which this model potential can be evaluated computationally allows us to average over a sufficiently large number of runs starting from different initial geometries and with different random number sequences to be able to report statistically relevant numbers $N_{\rm av}$ on the efficiency for a wide range of cluster and GA population sizes. Last, not least, despite the simplicity of the LJ potential the resulting PESs for different cluster sizes range from easy to quite hard with respect to the global optimization problem \cite{doye99}. This enables a systematic analysis of the algorithm performance not only over the relevant cluster size range, but also for both simple and complex PES topographies. Specifically, we therefore focus on the series of single-funnel LJ clusters with 10, 15, 20, 25, $\ldots$, 65, and 70 atoms for the prior aspect and contrast this with an investigation of the double-funnel LJ$_{38}$ cluster for the latter. The random structures to initialize the sampling runs are created by randomly placing atoms inside a sphere of radius 5 with the constraint that the distance between any pairs of atoms is more than 0.5. The reported values for $N_{\rm av}$ are then obtained by averaging over many runs starting from different random structures and random number seeds. The targeted convergence of $N_{\rm av}$ within 5 percent was usually achieved after about 100 different runs, and only for the largest clusters an averaging over up to 175 runs was required. For the local geometry optimization ensuing each move we employ the FIRE algorithm \cite{bitzek06} and consider the structure relaxed when all forces fall below $10^{-3}$. Correspondingly optimized cluster structures are identified as belonging to the same local PES minimum, if their relative energies differ by less than $10^{-3}$.

\section{Results}

\begin{figure}
\begin{center}
\scalebox{0.3}{\includegraphics[angle=270]{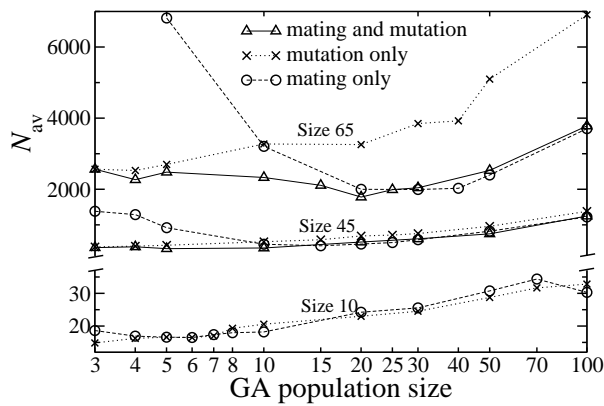}}
\end{center}
\caption{GA performance for the single-funnel LJ$_{10}$, LJ$_{45}$ and LJ$_{65}$ clusters. Shown is the dependence of the average number of trial structures $N_{\rm av}$ required to find the global minimum on the employed population size. The different lines for each cluster size correspond to GAs employing only mating moves, only mutation moves or both.
\label{fig1}}
\end{figure}

With equivalent results obtained for all single-funnel clusters, Fig. \ref{fig1} summarizes the performance data for the selected clusters LJ$_{10}$, LJ$_{45}$ and LJ$_{65}$ spanning the studied size range. Separately analyzed is the efficiency of the algorithm when allowing for only mutation or only mating moves, as well as for both. Intriguingly, it is primarily in the latter case that the algorithm exhibits an unusual robustness in the sense that the overall performance varies over less than a factor of two for a wide range of population sizes and for all cluster sizes studied. Recalling that the essential idea to employ a population was to potentially combine information from disparate parts of the high-dimensional PES through the mating of different cluster configurations, the large influence of the twinning mutations obtained particularly for the larger cluster sizes, e.g. LJ$_{65}$ in Fig. \ref{fig1}, is counterintuitive at first sight. In order to rationalize it, let us first further qualify what is meant with disparate parts. ``Disparate'' in this context does not exclusively refer to the PES topography, but also to the move types used. Any area of the PES is ``disparate'' from a current configuration, if it cannot be efficiently reached by the employed moves, and areas that are ``disparate'' for one move type may actually be quite ``close'' for another. The almost equal performance of the GA when using either only mutation or additionally mating moves for the smallest clusters like LJ$_{10}$ thus reflects that already the mutation moves alone enable quite efficient jumps everywhere on the corresponding PES of still rather limited dimensionality. In this situation there is no added value when enabling a mating of different members of the population. On the contrary, a larger population creates an increasing overhead as it takes more trial moves until the loop over all population members is completed. Especially when only enabling mutation moves, a larger population then primarily means that for a given number of trial moves every single cluster configuration is less often subject to a change  and the algorithm needs overall a larger $N_{\rm av}$ to more or less optimize all population members in parallel. Correspondingly and as illustrated by Fig. \ref{fig1}, for all cluster sizes the performance of a GA based exclusively on mutation moves decreases monotonously with population size. 

It is only for the larger cluster sizes that mating different population members starts to pay off, cf. Fig. \ref{fig1}. In the increasingly higher-dimensional PESs these moves are now apparently able to reach areas that can not be accessed that efficiently by mutations anymore. In the sense of genetic diversity this feature will be the better the larger the population. On the other hand, also here the increasing overhead argument applies, rationalizing why the performance curves for the mating-only GA in Fig. \ref{fig1} start to develop a minimum at an optimum population size that gets more pronounced the larger the cluster size. From the simplistic perspective of an increasing value brought about by mating in increasingly higher-dimensional PESs, the shift of this minimum to larger optimum population sizes for the larger clusters visible in Fig. \ref{fig1} is also quite intuitive. 

\begin{figure}
\begin{center}
\scalebox{0.3}{\includegraphics[angle=270]{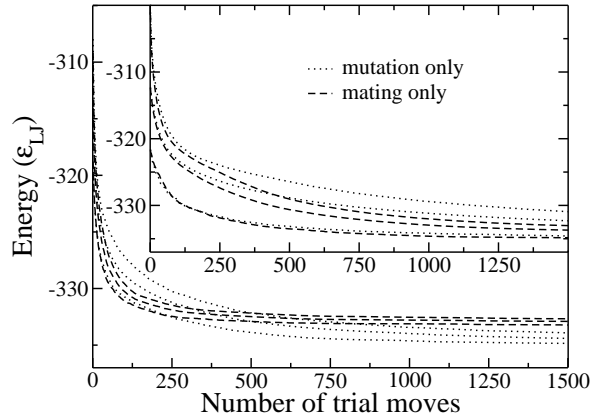}}
\end{center}
\caption{Energy of the GA population members in the sampling of LJ$_{65}$ as a function of the number of trial moves for a population of 5 (main plot) and 30 (inset). On each plot the three curves of the same type (dotted lines for mutation-only GA, and dashed lines for mating-only GA) represent the lowest, average and highest energy in the population.
\label{fig2}}
\end{figure}

While this view furthermore rationalizes the poor performance of mating-only GA at very small population sizes, the then much higher efficiency of the GA based solely on mutations is quite surprising. Analyzing the energies of the most and least favorable population member, as well as the average energy within the population during the sampling run we can nevertheless trace it back to limitations of small-population mating-only GA in bringing the system down through the lowest part of the PES funnel. Figure \ref{fig2} illustrates this for the LJ$_{65}$ cluster for both a very small population of 5, for which mating-only is much less efficient than mutation-only GA, and a larger population of 30, for which mating-only is more efficient than mutation-only GA, cf. Fig. \ref{fig1}. For the small population, the energy of the most stable cluster in the population decreases in both algorithm variants almost equally quickly in the initial 250 trial moves. At the same time the wider jumps in configuration space enabled by mating moves eliminate much more effectively the highest-energy members in the population and the average energy in the mating-only GA is brought down more rapidly than in the mutation-only variant. However, exactly this tendency of mating moves to induce more drastic changes to the cluster structure turns into a disadvantage in the continued sampling run, when in particular for the close-packed LJ clusters often only minor modifications to e.g. erase remaining dislocations or coordination faults are required to bring the system through the end of the PES funnel towards the global minimum. Correspondingly, Fig. \ref{fig2} shows a saturation of the mating-only GA run at higher energies than the mutation-only GA, where in the latter e.g. twinning mutations with small rotational angle can efficiently induce possibly required small structural modifications. This limitation is lifted in larger populations which now contain sufficiently many and diverse configurations that also mating moves inducing smaller structural modifications occur with sufficient probability. This is reflected in Fig. \ref{fig2}, where the lowest-energy curves for the population of 30 decrease now similarly for both algorithm variants, and the mating-only GA ultimately results as the more efficient approach. 

We therefore arrive at an interesting performance crossover between mating-only and mutation-only GA as a function of population size. For large and therewith diverse enough populations the added value brought about by mating kicks in and mating-only GA is preferred. On the contrary, if the population is too small this diversity bonus no longer applies. While still not as efficient as mating-only GA at optimum population size, mutation-only GA is then the more robust approach as it is better able to cope with the possible requirement of subtle structural modifications when already close to the optimum geometry. Most intriguingly, yet with the presented understanding also quite naturally, Fig. \ref{fig1} shows that these two effects are to some degree additive in the sense that the GA implementation allowing both mutation and mating exhibits a performance that in the two extremes of very small and very large populations is virtually identical to the performance of mutation-only and mating-only GA, respectively. In the end, this leads to an efficiency that is very robust with respect to the employed population size for all cluster sizes studied and varies by less than a factor of two for the wide range of populations shown in Fig. \ref{fig1}.

\begin{figure}
\begin{center}
\scalebox{0.33}{\includegraphics[angle=0]{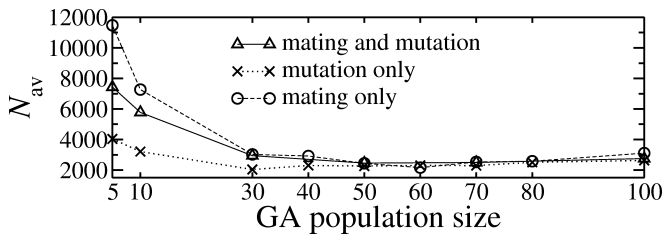}}
\end{center}
\caption{Same as Fig.~\ref{fig1}, but for the double-funnel LJ$_{38}$ cluster.
\label{fig3}}
\end{figure}

This result largely carries over to the more complex PES topography of the double-funnel LJ$_{38}$ cluster. As shown in Fig. \ref{fig3}, also here the mutation-only GA is most efficient at small population sizes. That this efficiency is in this case so pronounced compared to the mating-only GA nicely corroborates the interpretation developed from the single-funnel cluster data. As discussed in detail by Wolf and Landman \cite{wolf98}, there are a multitude of structures in the energetic neighborhood of the LJ$_{38}$ global minimum, a fcc truncated octahedron, that essentially correspond to multitwinned fcc crystallites. With the initial trial moves likely to bring the system into one of these low-energy configurations, it is particularly twinning mutation moves that can then very efficiently reach the optimum pure fcc structure. In the present GA implementation applying both mating and mutation, these twinning mutations do not occur too often in the loop over all pairings, which is why the performance of this variant in this regime is still worse than mutation-only, albeit significantly improved compared to mating-only GA. Obviously, this suggests that the actual degree to which mutation moves are admixed may be optimized for specific systems, yet what we consider the more important conclusion from our work is that whatever the percentage, mutation admixture seems to generally improve the robustness of the GA algorithm with respect to the employed population size. From the presented understanding of enabling small structural modifications we believe that this feature is in fact not restricted to the actual twinning mutations used here, but should equally hold for e.g. single or collective particle moves in which one or several atoms of the cluster are randomly displaced, as well when using more sophisticated crossover operations.

\section{Conclusions}

In conclusion, we have investigated the performance dependence of a ``cut and splice'' genetic algorithm on the employed population size. Focusing on the geometry optimization of both a series of single-funnel LJ clusters up to 70 atoms and the double-funnel LJ$_{38}$ cluster we generally obtain that admixing twinning mutation moves leads to an improved robustness of the algorithm. Here, improved robustness denotes an efficiency in finding the global minimum that varies less over a wide range of population sizes studied and that in particular for small populations is significantly improved compared to a mating-only GA. The latter is traced back to an enhanced possibility of small structural modifications offered by mutation moves. Especially for systems favoring close-packed structures like the LJ clusters this is essential to erase remaining dislocations or coordination faults and to bring the system through the end of the PES funnel towards the global minimum. Within this understanding we believe that this finding is not specific to LJ interactions, nor is it restricted to the ``cut and splice'' mutation and mating moves studied here. This suggests that is is in general recommended to admix mutation moves, not only because of their well-known beneficial effect on the sampling \cite{wolf98}, but also to increase the robustness of the GA algorithm with respect to the {\em a priori} unknown parameter population size. 

The latter feature is particularly important for first-principles GA sampling, and in this respect the very stable performance of the present mating+mutation GA implementation, achieved without any system-specific tuning, is very promising. To set this into perspective, one should compare the about 2000 moves required to find the ground state of the complex LJ$_{38}$ PES with any reasonable population size to the roughly equal number of moves needed by the classic basin-hopping (BH) scheme \cite{wales99}, when its central technical parameter, the effective temperature, is specifically optimized to the double-funnel PES topography \cite{rossi09}. Latest reports indicate that more sophisticated GA \cite{schoenborn09} or BH \cite{rossi09} schemes (including the recently introduced minima-hopping \cite{goedecker04}) may reduce this number by a factor $\sim 2$, but at the expense of introducing a manifold of technical parameters. While not large, such a reduction would still be highly desirable in view of the high computational cost incurred by first-principles based sampling. Not withstanding, it would not be of much use if only achieved after excessive parameter optimization for a specific PES landscape. Consequently, we believe that further progress in this direction requires revisiting the global optimization problem from the perspective of algorithmic robustness as done in the present work.

\section{Acknowledgments}

Funding within the MPG Innovation Initiative ``Multiscale Materials Modeling of Condensed Matter'' is gratefully acknowledged.


\begin{thebibliography}{99}

\bibitem{jena87}
Special Feature on {\em Cluster Chemistry and Dynamics}, Proc. Natl. Acad. Sci. U.S.A. {\bf 103 (28)} (2006).

\bibitem{wales03}
D.J. Wales, {\em Energy Landscapes}, Cambridge University Press, Cambridge (2003).

\bibitem{goldberg89}
D.E. Goldberg, {\em Genetic Algorithms in Search, Optimization, and Machine Learning},
Kluwer Academic Publishers, Boston (1989).

\bibitem{hartke04}
B. Hartke, Struct. and Bonding {\bf 110}, 33 (2004).

\bibitem{deaven95} 
D.M. Deaven and K.M. Ho, Phys. Rev. Lett. {\bf 75}, 288  (1995).

\bibitem{deaven96}
D.M. Deaven, N. Tit, J.R. Morris, and K.M. Ho, Chem. Phys. Lett. {\bf 256}, 195 (1996).

\bibitem{wolf98} 
M.D. Wolf and U. Landman, 
J. Phys. Chem. A {\bf 102}, 6129 (1998). 

\bibitem{hartke99}
B. Hartke, J. Comput. Chem. {\bf 20}, 1752 (1999).

\bibitem{johnston03}
R.L. Johnston, Dalton Transactions 4193 (2003).

\bibitem{wales97}
D.J. Wales and J.P.K. Doye, 
J. Phys. Chem. A {\bf 101}, 5111 (1997). 

\bibitem{cambridge_database} 
D.J. Wales {\em et al.}, 
{\em  The Cambridge Cluster Database}.

\bibitem{doye99}
J.P.K. Doye, M.A. Miller, and D.J. Wales, J. Chem. Phys. {\bf 111}, 8417 (1999).

\bibitem{bitzek06} 
E. Bitzek {\em et al.},
Phys. Rev. Lett. {\bf 97}, 170201 (2006).

\bibitem{wales99}
D.J. Wales and H.A. Scheraga, Science {\bf 285}, 1368 (1999).
 
\bibitem{rossi09}
G. Rossi and R. Ferrando, J. Phys.: Condens. Matter {\bf 21}, 084208 (2009).

\bibitem{schoenborn09}
S.E. Schoenborn, S. Goedecker, S. Roy, and A.R. Oganov, J. Chem. Phys. ({\em in press});
arXiv:0810.2055.

\bibitem{goedecker04}
S. Goedecker, J. Chem. Phys. {\bf 120}, 9911 (2004).

\end{thebibliography}
\end{document}